\documentstyle[12pt]{article}

\renewcommand{\thefootnote}{\fnsymbol{footnote}}

\newcommand{\newsection}{    % Numeration of eqs. is automatic
\setcounter{equation}{0}
\section}
\newcommand{\non}{\nonumber \\*}

\newcommand{\eq}[1]{Eq.~(\ref{#1})}
\def \ov {\over }
\def\bea{\begin{eqnarray}}
\def\eea{\end{eqnarray}}

\def \de{\partial}

\def\LB{\left(}
\def\RB{\right)}
\def\be{\begin{equation}}
\def\ee{\end{equation}}
\def\la{\label}
\def \bi{\bibitem}

\def\LA{\langle}
\def\RA{\rangle}
\def\a{\alpha}
\def\b{\beta}
\def\t{\theta}
\def\T{\Theta}
\def\d{\delta}
\def\g{\gamma}
\def\de{\partial}
\def\ep{\epsilon}
\def\G{\Gamma}
\def\ex{{\rm exp}}
\def\lde{{\stackrel{\leftarrow }{\partial }}}
\def\rde{{\stackrel{\rightarrow }{\partial }}}
\def\br{\mbox{\boldmath$\rho$}}
\def\bT{\mbox{\boldmath$\Theta$}}
\def\bxi{\mbox{\boldmath$\xi$}}
\def\bX{\mbox{\boldmath$X$}}

\def\bH{\mbox{\boldmath${\cal H}$}}
\def\r{\rho}
\def\e{{\,\rm e}\,}
\def\L{{\cal L}}
\def\D{{\cal D}}

\font\mybb=msbm10 at 12pt
\def\bb#1{\hbox{\mybb#1}}
\def\B {\bb{B}}
\def\C{\bb{C}}
\def\F{\bb{F}}
\def\LA{\langle}
\def\RA{\rangle}
\def\gh{{\rm gh}}
\def\vep{\varepsilon}
\def\z{\zeta}
\def\BV{{\hbox{\footnotesize{\rm{BV}}}} }
\def\inte{{\hbox{\footnotesize{\rm int}}}}
\def\H{{\cal H}}

\begin{document}

\setcounter{page}{1}
\renewcommand{\thefootnote}{\arabic{footnote}}
\setcounter{footnote}{0}

\begin{titlepage}
\begin{flushright}
CALT-68-2434\\
\end{flushright}
\vspace{.5cm}

\begin{center}
{\LARGE  A Path Integral Approach To Noncommutative Superspace}\\
\vspace{1.1cm}
{\large Iouri Chepelev${}$\footnote{
E-mail: chepelev@theory.caltech.edu } 
and  Calin Ciocarlie${}$\footnote{ \ E-mail: calin@theory.caltech.edu } }\\
\vspace{18pt}

{\it Department of Physics}

{\it California Institute of Technology}

{\it  Pasadena, CA 91125, USA}
\\
\end{center}
\vskip 0.6 cm

\begin{abstract}
A path integral formula for
the associative star-product of two superfields is proposed. 
It is a generalization of the Kontsevich-Cattaneo-Felder's formula for
the star-product of functions of bosonic coordinates. 
The associativity of the star-product imposes certain conditions on the background of  our sigma model.
 For generic background the 
action is not supersymmetric. The supersymmetry invariance of
the action constrains the background and leads to a simple formula
for the star-product. 

\end{abstract}

\end{titlepage}

\setcounter{footnote}{0}

\newsection{Introduction}

An outstanding problem is the following:  {\it Given a classical structure find its quantum extension(s)}. Deformation quantization theory is an interesting 
approach to this problem\cite{BFFLS}. 

In \cite{wilde,fedosov} it was shown that any symplectic manifold can be deformation-quantized.
The question whether any Poisson manifold can be quantized was 
answered in affirmative only recently by Kontsevich \cite{kont}. The solution presented
by Kontsevich uses in a essential way ideas of perturbative string theory.
Kontsevich's construction was further clarified by Cattaneo and Felder  in \cite{CF}, where an explicit
path integral
formula for the star-product
of functions on  the Poisson manifolds was given. 

Unlike the star-product on the Grassmann-even manifolds, the star-product on the supermanifolds is not sufficiently studied. For a list of
papers on the subject, see \cite{superstar,NSUSY}. 
Ref. \cite{NSUSY} gives a systematic analysis of 
the deformation of the algebra of four dimensional supercoordinates.\footnote{
For an earlier discussion of the quantum deformations of the Poincare supergroup, see
\cite{kosinsky}.} 
The most general deformed $N=1$, $d=4$ supercoordinate algebra which satisfies Jacobi identities and is compatible with the 
$N=1$ supersymmetry reads
\be
[x^i,x^j] = i \hbar ~\B^{ij}(\t),~~~~~[x^i,\t^{\a}] = i \hbar ~\C^{\a i},~~~~~\{ \t^{\a},\t^{\b}\} = 0,
\la{susyalg}
\ee
where $\B$ is $x$-independent and is a linear function of $\t$. $\C$ is independent of $x$ and $\t$, and satisfies
an equation following from the Jacobi identity.

Recently, using Berkovits' covariant formalism for quantizing type II superstring compactified on a Calabi-Yau
three-fold \cite{berkovits}, Ooguri and Vafa made an interesting observation that the RR field strengths can give
rise to a non-zero $\{ \t, \t \}$ anticommutator  in four dimensions \cite{OV}.\footnote{For an earlier
discussion of RR backgrounds in the context of noncommutativity in string theory, see \cite{schiappa}.}  
De Boer, Grassi and van Nieuwenhuizen \cite{van} generalized this observation to ten dimensional superspace using the
covariant formalism \cite{berkovits2,covan}. Ref.\cite{van} also pointed out that the gravitino 
yields a non-zero value for $[x,\t]$.

The purpose of this paper is to give a general path integral formula for the star-product of functions of $x$ and $\t$.
The $\star$-algebra which arises from our path integral formula is of the form 
$$
[x^i,x^j]_{\star} = i \hbar  ~\B^{ij}(x,\t) + O(\hbar^2),
$$
\be
[x^i,\t^{\a}]_{\star} = i \hbar  ~\C^{\a i}(x,\t)+O(\hbar^2),~~~~~~
\{ \t^{\a},\t^{\b}\}_{\star} = i \hbar ~\F^{\a\b}(x,\t) + O(\hbar^2).
\la{susyalgmod}
\ee
Using path integral methods, we will derive  the associativity equations satisfied by the fields $\B$, $\C$ and $\F$.

As a main tool in deriving the path integral formula, we use an elegant supermanifold
formalism for the Batalin-Vilkovisky(BV) quantization \cite{BV1}
introduced by  Alexandrov, Kontsevich, Schwarz and Zaboronsky\cite{AKSZ}. This formalism was
 further elaborated and extended by Park\cite{jaesuk}.\footnote{
See also \cite{CF2}.}
The AKSZ method \cite{AKSZ} is a method
to construct solutions of the BV master equation directly, without starting from a classical action with a 
set of symmetries, as is done in the BV method. The classical action is then recovered {\it a posteriori} by
setting the fields of non-zero ghost number to zero. In ten spacetime dimensions
 the classical action that we obtain from our BV-action
using this method will turn out to be equivalent, for the constant background fields, to the one used in 
\cite{van}.

The star-product will be
represented as a path integral of a sigma model on a superdisk.
 The BV-action $S=S[\bX,\bT,\br,\bxi]$
of the sigma model will be  a functional of  the superspace-valued
superfields. The action $S$ depends on the background fields $\B$, $\C$ and $\F$. Imposing 
the supersymmetry invariance of the action, we will obtain the algebra (\ref{susyalg}). We will also find
an explicit formula for the supersymmetric star-product of functions of $x$ and $\t$.

The paper is organized as follows. In section 2  the path integral formula is given, the associativity
equations are derived and the Feynman diagrammatics is presented. The supersymmetric backgrounds
are considered in section 3 where an explicit expression for the star-product is given. A summary of our results
 can be found in section 4.

\newsection{Path integral formula} 
The star-product is
represented as a path integral on a superdisk with the two functions inserted on the boundary of
the disk: 
$$
f\star g~(x,\t) = 
$$
\be
 \int_{X(\infty)=x, \T(\infty)=\t} f(X(0),\T(0)) g(X(1),\T(1))~ \e^{ {i\ov \hbar} S[ \hbox{\footnotesize{\bX}},\hbox{\footnotesize{\bT}},\hbox{\footnotesize{\br}},\hbox{\footnotesize{\bxi}}]} ~D\bX D\bT D\br D\bxi
\bigg|_{{\footnotesize{\rm Lagrangian}\atop
\footnotesize{\rm submanifold}}}
\la{formula}
\ee
where $0$ and $1$ are the points on the boundary of the disk where the functions are inserted. The path
integral is computed with the condition $X=x$, $\T=\t$ at the point $\infty$ of the boundary of the
disk. The path integral is taken over a Lagrangian submanifold in the field space $\{ \bX, \bT, \br, \bxi\}$.

We first describe the fields that appear in  \eq{formula}.
$\bX$, $\bT$, $\br$ and $\bxi$ are maps from the superdisk to a superspace ${\cal M}$.
The local coordinates of ${\cal M}$ are $X^i, \T^{\a}, \rho_i$ and $\xi_{\a}$. For simplicity we work in the spacetime dimension $n$ that admits a Majorana representation. $\T$ is a Majorana spinor in $n$ dimensions.
 $i$ is a spacetime index
and $\a$ is a spinorial index. 
The coordinates of ${\cal M}$
are  graded with respect to two Grassmann gradations $\vep_1$ and $\vep_2$:
\bea
&&\vep_1 (X) = 0,~~~~~~\vep_1 (\T)=1,~~~~~~\vep_1(\rho)=0,~~~~~~\vep_1(\xi)=1, \non
&&\vep_2 (X) = 0,~~~~~~\vep_2 (\T)=0,~~~~~~\vep_2(\rho)=1,~~~~~~\vep_2(\xi)=1.
\la{parity}
\eea
The commutation property of functions on ${\cal M}$ is given by the relation 
\be
f g = (-1)^{\vep_1(f)\vep_1(g)+\vep_2(f)\vep_2(g)} g f .
\ee
Let  $\sigma^{\mu}$ and $\zeta^{\mu}$ ($\mu =1,2$) be bosonic and fermionic coordinates of the
superdisk.
Then
\be
\bX = \bX(\sigma,\zeta),~~~\bT = \bT(\sigma,\zeta),~~~\br =\br(\sigma,\zeta),~~~\bxi = \bxi(\sigma,\zeta).
\ee
In other words  we have an embedding $\phi$ of the superdisk  $D$ into the superspace ${\cal M}$:
\be
\phi : D \rightarrow {\cal M}.
\la{embed}
\ee
We now describe the action $S$ in \eq{formula}. It is an integral over the superdisk $D$ 
\be
S= S_0 + S_{\hbox{\footnotesize{\rm int}}} = \int_D  d^2\sigma d^2\zeta  ~(\L_0 +\bH) =  \int_D  d^2\sigma d^2\zeta ~\L 
\la{action}
\ee
of the Lagrangian density
\bea
\L &=& \L_0 + \bH,  \non
\L_0 &=& \br_i D\bX^i + \bxi_{\a} D\bT^{\a}, \non
\bH&=& {1\ov 2} \B^{i j}(\bX,\bT) \br_{i}\br_{j} + {1\ov 2} \F^{\a\b}(\bX,\bT)\bxi_{\a} \bxi_{\b}+
\C^{\a i}(\bX,\bT) \bxi_{\a}\br_{i} .
\la{Lagrangian}
\eea
$D$ in \eq{Lagrangian} is defined as
\be
D= \zeta^{\mu} {\de\ov \de \sigma^{\mu}}.
\ee
The action $S$ is a Batalin-Vilkovisky (BV) action. The components of the superfields 
$\bX$, $\bT$, $\br$ and $\bxi$ in the expansion in $\zeta$,
\bea
\bxi &=& \xi^{(0)} ~+~ \xi^{(1)}_{\mu}\z^{\mu} ~+~ \xi^{(2)}_{\mu\nu} \z^{\mu}\z^{\nu}, \non
\br &=& \rho^{(0)} ~+~ \rho^{(1)}_{\mu}\z^{\mu} ~+~ \rho^{(2)}_{\mu\nu} \z^{\mu}\z^{\nu},\non  
\bT &=& \T^{(0)} ~+~ \T^{(1)}_{\mu}\z^{\mu} ~+~ \T^{(2)}_{\mu\nu} \z^{\mu}\z^{\nu}, \non
\bX &=& X^{(0)} ~+~ X^{(1)}_{\mu}\z^{\mu} ~+~ X^{(2)}_{\mu\nu} \z^{\mu}\z^{\nu}, 
\la{compon}
\eea
can be identified with the 
fields and antifields in the BV quantization of a classical action $S_{cl}$. The latter can be obtained from 
\eq{action} setting to zero fields with non-zero ghost number.
The ghost numbers of the component fields in \eq{compon} can be computed by assigning ghost number one
to $\zeta$ and taking overall ghost number of the superfield to be given by the parity $\vep_2$ defined
in \eq{parity}. Thus we find
\be
\gh (X^{(k)}) =  -k,~~\gh (\T^{(k)})= -k,~~\gh (\rho^{(k)})=1-k,~~\gh (\xi^{(k)})=1-k,~~~~k=0,1,2.
\ee
Note that $\bH$ is the most general ghost number  two interaction term that can be constructed out of
$\bX$, $\bT$, $\br$ and $\bxi$.
Setting to zero fields with a non-zero ghost number in \eq{action}, we find
\bea
 S_{cl}&=&\int ~[\rho^{(1)}_i dX^{(0)i} + \xi^{(0)}_{\a} d\T^{(0)\a}+
{1\ov 2} \B^{i j}(X^{(0)},\T^{(0)}) \rho^{(1)}_{i}\rho^{(1)}_{j}\non
&&~~~~~~~~ + {1\ov 2} \F^{\a\b}(X^{(0)},\T^{(0)})\xi^{(1)}_{\a} \xi^{(1)}_{\b}+
\C^{\a i}(X^{(0)},\T^{(0)}) \xi^{(1)}_{\a}\rho^{(1)}_{i}  ],
\eea
where 
\be
\rho^{(1)}_i = \rho^{(1)}_{i\mu}d\sigma^{\mu},~~~~~\xi^{(1)}_{\a} = \xi^{(1)}_{\a\mu}d\sigma^{\mu}
\ee
are one forms.\footnote{Integrating out $\rho^{(1)}$ in $S_{cl}$, we end up with an action which is, for
constant background fields,
equivalent to the action (without the $G \de X\de X$ term) used in \cite{van}. To be more precise, the two actions are
equivalent modulo the difference in the type of spinors considered: we have type IIA spinors whereas ref.\cite{van}
used type IIB.} 

In the BV quantization the action $S$ must satisfy the master equation \cite{BV1}:
\be
(S,S)_{\hbox{\footnotesize{\rm{BV}}}}- 2 i\hbar \Delta S = 0,
\ee
where $(\cdot,\cdot)_{\BV}$ is the BV-bracket and $\Delta$ is the BV-Laplacian. 
It can be checked that\cite{CF} 
\be
\Delta S = 0.
\ee
Thus the master equation becomes
\be
(S,S)_{\hbox{\footnotesize{\rm{BV}}}} = 0.  
\la{SS}
\ee
The associativity of the star-product follows from
the Ward identity as in \cite{CF}. Thus \eq{SS} implies the associativity of the star-product:
\be
(S,S)_{\hbox{\footnotesize{\rm{BV}}}} = 0  ~~~~~~\Longrightarrow~~~~~~~  \hbox{{\rm ~the~star-product~is~associative}}. 
\ee  

We decompose the action $S$ into the kinetic $S_0$ and the interaction $S_{\inte}$ parts as in \eq{action}.
 It is easy to show that (see \cite{jaesuk}):
\be
(S_0,\bX)_{\BV}=D\br,~~~(S_0,\br)_{\BV}=D\bX,~~~(S_0,\bT)_{\BV}=D\bxi,~~~(S_0,\bxi)_{\BV}=D\bT .
\ee
Thus we have
\be
(S_0,S_{\inte})_{\BV}=\int_D  D \bH .
\ee
Since the integrand on the r.h.s. of this equation is a total derivative, the integral can be reduced to
 an integral over the boundary of the disk.
 This boundary integral vanishes due to the boundary conditions for the superfields:
\bea
&&X^{(1)}_{\mu} n^{\mu} = 0,~~~X^{(2)}_{\mu\nu} = 0,~~~\rho^{(0)} = 0,~~~\rho^{(1)}_{\mu} t^{\mu}=0, \non
&&\T^{(1)}_{\mu} n^{\mu} = 0,~~~\T^{(2)}_{\mu\nu} = 0,~~~\xi^{(0)} = 0,~~~\xi^{(1)}_{\mu} t^{\mu}=0, 
\eea
where $n^{\mu}$ and $t^{\mu}$ are normal and tangential vectors to the boundary of the disk.\footnote{
The components $X^{(0)}$ and $\T^{(0)}$ satisfy the conditions: 
$
X^{(0)}(\infty)= x$ and $\T^{(0)}(\infty) = \t
$.}
Thus we have $(S_0,S_{\inte})_{\BV}=0$. Similarly, we can show that
$(S_0,S_0)_{\BV}=0$.
Thus \eq{SS} reduces to
 \be
(S_{\inte},S_{\inte})_{\BV}=0.
\la{sint}
\ee

\eq{sint}
implies that 
the background fields $\B$, $\F$ and $\C$ satisfy certain equations. We now derive these
 equations. 
The calculations can be significantly simplified if we use the following formula\cite{AKSZ,jaesuk} 
 which relates the BV-bracket
to a bracket $[\cdot , \cdot ]$ on the superspace ${\cal M}$:
\be
\int_D \phi^* \LB [f,g] \RB = \LB \int_D  \phi^* f, \int_D \phi^* g \RB_{\hbox{\footnotesize{\rm BV}}}. 
\la{parkid}
\ee
Here $\phi^* f$ and $\phi^* g$ are pullbacks of the functions $f$ and $g$ on ${\cal M}$ to the
superdisk $D$ with respect to the map $\phi$ from \eq{embed}.
The bracket $[f,g]$  is given by
\be
[f,g] = {\de_r f\ov \de \rho_i} {\de g\ov \de X^i} -  {\de f\ov \de X^i}{\de_l g\ov \de \rho_i}+{\de_r f\ov \de \xi_{\a}} {\de_l g\ov \de \T^{\a}} + {\de_r f\ov \de \T^{\a}} {\de_l g\ov \de \xi_{\a}}.
\la{bracket}
\ee
where $r$ and $l$ denote the right and the left partial derivatives.\footnote{\eq{bracket} follows from
a general formula for the (odd) Poisson bracket on the supermanifold ${\cal M}$ \cite{AKSZ}. Let $z^a$ be the local coordinates of ${\cal M}$ and let
$\omega = dz^a \omega_{ab} dz^b$ be an odd non-degenerate closed 2-form on ${\cal M}$. In our case it is
given by $\omega = 2 (d X^i \wedge d \rho_i + d\T^{\a}\wedge d\xi_{\a})$. 
The Poisson bracket is given by
$[f,g] = {\de_r f\ov \de z^a }\omega^{ab}{\de_l g\ov \de z^b}$,
where
 $\omega^{ab}$ is the inverse matrix of $\omega_{ab}$.
}
Using \eq{parkid}, \eq{sint}  is equivalent to
\be
[\H, \H ] = 0,  
\la{HH}
\ee
where $\H$ is a function on ${\cal M}$:
\be
{\cal H}= {1\ov 2} \B^{i j}(X,\T) \r_{i}\r_{j} + {1\ov 2} \F^{\a\b}(X,\T)\xi_{\a} \xi_{\b}+
\C^{\a i}(X,\T) \xi_{\a}\r_{i}.
\la{Hfunction}
\ee
Since $\vep_1(\H) = \vep_2(\H) = 0$, we have
\be
[\H,\H]= 2 \LB {\de_r \H\ov \de \rho_i} {\de \H\ov \de X^i} + {\de \H\ov\de\xi_{\a}}{\de_l\H\ov\de\T^{\a}}\RB .
\la{HH1}
\ee
From the equations (\ref{HH}), (\ref{Hfunction}) and \eq{HH1} we find
\bea
(\C^{\a i} \de_{\a} \B^{j k}+ \B^{il}\de_l \B^{jk})  ~~\rho_i \rho_j \rho_k &=& 0, \non
(\F^{\a\d} \de_{\d} \F^{\b\g}+ \C^{\a i}\de_i \F^{\b\g})~~\xi_{\a}\xi_{\b}\xi_{\g} &=& 0, \non
( \F^{\a\b}\de_{\b}\B^{i j} - 2 \C^{\b i}\de_{\b} \C^{\a j}-2\B^{ik}\de_k \C^{\a j}+\C^{\a k}\de_k \B^{ij})~~\xi_{\a} \rho_i \rho_j &=& 0, \non
(2 \F^{\a\d} \de_{\d} \C^{\b k} - \C^{\d k}\de_{\d} \F^{\a\b}+2 \C^{\a i}\de_i \C^{\b k}- \B^{k j}\de_j \F^{\a\b})~~ \xi_{\a}\xi_{\b} \rho_k &=& 0, 
\la{master}
\eea
where $\de_i=\de / \de X^i$ and $\de_{\a}=\de_l /\de \T^{\a}$. \eq{master} is a generalization
of the familiar Jacobi identity for the Poisson bivector $B^{ij}$:
\be
B^{il}\de_l B^{jk}+ B^{jl}\de_l B^{ki}+B^{kl}\de_l B^{ij}=0.
\ee

We now explain the last ingredient of the formula (\ref{formula}): the integration over a Lagrangian submanifold
in the field space $\{ \bX, \bT, \br, \bxi\}$. A final step in the BV quantization procedure involves choosing a ghost number $-1$ function $\Psi$,
the "gauge fixing fermion", of the fields\cite{BV1}. The path integral is then taken over
the Lagrangian submanifold defined by the equations:
\be
{\rm antifield}={\de \Psi\ov \de \hbox{{\rm field}}}
\ee
We refer the reader to ref.\cite{CF} for the details on how new fields, called antighosts and Lagrange multipliers
together with their antifields, are added to the set of fields $\bX$ and $\br$. In a very similar way, we can add
antighosts and Lagrange multipliers
together with their antifields to $\bT$ and $\bxi$. This procedure results in the following operational formula  for  the star-product:
\bea
&&f\star  g~(x,\t) = \bigg\langle f(x+\bX(0,0),\t+\bT(0,0))\non
&&~~~~~~\e^{{i\ov \hbar}S_{\hbox{\inte}}[ x+\hbox{\footnotesize{\bX}},\t+\hbox{\footnotesize{\bT}},\hbox{\footnotesize{\br}},\hbox{\footnotesize{\bxi}}]}g(x+\bX(1,0),\t+\bT(1,0))\bigg\rangle ,
\la{formula2}
\eea
where $S_{\inte}$ is given by \eq{action} and \eq{Lagrangian}.\footnote{The tadpole diagrams
in \eq{formula2} are discarded. This can be achieved by the introduction of terms linear in $\br$ and $\bxi$ in
the action: $\de_i \B^{ij}\br_j$, $\de_{\a}\F^{\a\b}\bxi_{\b}$, $\de_{\a}\C^{\a i}\br_i$ and $\de_i \C^{\a i}\bxi_{\a}$.} 
The Wick-contraction $\langle \cdots \rangle$ in \eq{formula2} is done using the super-propagators:
\bea
\LA \bX^k(z_1,\zeta_1) \br_j(z_2,\zeta_2) \RA &=& {i \hbar \ov 2\pi} \d^k_j  \D\phi(z_2,z_1),\non
\LA \bT^{\a}(z_1,\zeta_1) \bxi_j(z_2,\zeta_2) \RA &=& {i \hbar \ov 2\pi} \d^k_j  \D\phi(z_2,z_1),
\la{propag}
\eea
where
\be
\phi (z,w)= {1\ov 2 i} {\rm ln} {(z-w)(z-{\bar w})\ov ({\bar z}-{\bar w})({\bar z}-w)}
\ee
and
\be
\D = \zeta_1^{\mu}{\de \ov \de z_1^{\mu}} + \zeta_2^{\mu} {\de \ov \de z_2^{\mu}}.
\ee
Note that points $z_1$ and $z_2$ from \eq{propag} belong to the upper-half complex plane. 
A generic term in the perturbative expansion in \eq{formula2} is of the form
\bea
&&\int \prod dz_{j} d\zeta_{j}\prod dz^{\prime}_{j} d\zeta^{\prime}_{j}~~\bigg\langle \bX(z_{1},\zeta_{1})\cdots \bX(z_{N},\zeta_{N}) ~\br(z_{N+1},\zeta_{N+1})\cdots \br(z_{2N},\zeta_{2N})~\non
&&~~~~~~~~~~~~~~~~~~\bT(z^{\prime}_{1},\zeta^{\prime}_{1})\cdots \bT(z^{\prime}_{M},\zeta^{\prime}_{M}) ~\bxi(z^{\prime}_{M+1},\zeta^{\prime}_{M+1})\cdots \bxi(z^{\prime}_{2M},\zeta^{\prime}_{2M})
\bigg\rangle
\la{perturb}
\eea 
times an expression involving  partial derivatives of $f(x,\t)$, $g(x,\t)$, $\B(x,\t)$, $\F(x,\t)$ and $\C(x,\t)$.
Wick-contracting using \eq{propag} and integrating over the Grassmann variables $\zeta$'s in \eq{perturb}, we
end up with a sum of expressions of the form
\be
\int d\phi(z_{i_1},z_{j_1}) \wedge d\phi(z_{i_2},z_{j_2}) \cdots d\phi(z_{i_N},z_{j_N}) \wedge
d\phi(z^{\prime}_{i_1},z^{\prime}_{j_1}) \wedge d\phi(z^{\prime}_{i_2},z^{\prime}_{j_2}) \cdots 
d\phi(z^{\prime}_{i_M},z^{\prime}_{j_M}) 
\ee
where $d\phi$ is defined as
\be
d\phi(z,w) = dz {\de\ov \de z}\phi(z,w)+dw {\de \ov \de w}\phi(z,w).
\ee

\newsection{A special case: supersymmetric backgrounds}
The kinetic part $S_0$ of the action (\ref{action}) is invariant under the following supersymmetry
transformation:
\bea
\d \bT^{\a} &=& \ep^{\a}, \non
\d \bX^k &=& i {\bar \ep} \G^k \bT , \non
\d \bxi_{\a}  &=& - i \br_k  {\bar\ep}^{\b} \G^k_{\b\a}, \non
\d \br_k &=&  0.  
\la{susy}
\eea
Since $\T$ is a Majorana spinor, the  transformation (\ref{susy}) is an $N=1$ or $N=2$ supersymmetry
transformation  depending 
on the dimension $n$ of the spacetime. For $n=3,4,8,9$ it is $N=1$ and for $n=2,10$ it is
$N=2$.
Imposing the invariance of the full action (\ref{action})  under the transformation (\ref{susy}), we find that
the background fields are further constrained. We will shortly show that the associativity and 
supersymmetry constrain the background fields as follows:
\be
\F = 0,
\la{Ffield}
\ee
$\B$ is $X$-independent and is  linear in $\T$:
\be
\B^{ij}(\T) = B^{ij} + i \T^{\a} [ (\G^0\G^i)_{\a\b}\C^{\b j}- (\G^0\G^j)_{\a\b}\C^{\b i} ],
\la{Bfield}
\ee
$\C$ is constant and satisfies the following equation
\be
\sum_{\pi \in {\rm permutations~of~}i,j,k} (-1)^{\pi} \C^{\a \pi(i)} (\G^0 \G^{\pi(j)})_{\a\b} \C^{\b \pi(k)}  = 0.
\la{jacobi}
\ee
Thus we recover the result of \cite{NSUSY} on the deformation of $N=1$, $d=4$  supercoordinate algebra.
In our calculations we do not assume that the spacetime dimension is four.
Thus the form of the deformed supercoordinate algebra that we get from our star-product 
is the same in any spacetime dimension that admits a Majorana representation.

In the supersymmetric case we consider $x$-independent backgrounds. The action (\ref{action}) is supersymmetric if the background
fields satisfy the following equations
\bea
\de_{\g} \F^{\a\b}  ~\xi_{\a}\xi_{\b} &=& 0, \non
(\de_{\g} \C^{\b j} + i (\G^0 \G^j)_{\g \a} \F^{\a\b})~\xi_{\b}\rho_j  &=& 0, \non
(\de_{\a} \B^{ij} - 2 i (\G^0 \G^i)_{\a\b}\C^{\b j})~\rho_i\rho_j &=& 0.  
\la{susyeq}
\eea 
The first equation in \eq{susyeq} implies that $\F^{\a\b}$ does not depend on $\T$. 
For the $x$-independent backgrounds and the constant $\F$ the master equation (\ref{master})
reduces to
\bea
\C^{\a i} \de_{\a} \B^{j k} ~~\rho_i \rho_j \rho_k &=& 0, \non
( \F^{\a\b}\de_{\b}\B^{i j} - 2 \C^{\b i}\de_{\b} \C^{\a j})~~\xi_{\a} \rho_i \rho_j &=& 0, \non
 \F^{\a\d} \de_{\d} \C^{\b k} ~~ \xi_{\a}\xi_{\b} \rho_k &=& 0. 
\la{masterpr}
\eea
Since $\xi$'s commute, 
the third equation in \eq{masterpr} is equivalent to
\be
\F^{\a\d} \de_{\d}  \C^{\b k} + \F^{\b \d} \de_{\d} \C^{\a k} = 0.   
\la{feq}
\ee
Using the second equation from \eq{susyeq},   \eq{feq} can be rewritten as
\be
\F^{\a \g} (\G^0\G^k)_{\g\d} \F^{\d \b}+\F^{\b \g} (\G^0\G^k)_{\g\d} \F^{\d \a} = 0.
\ee
Setting $k=0$ in this equation, we have
\be
\F^{\a\mu} \d_{\mu\nu} \F^{\nu \b}=0.
\ee
Since the action (\ref{action}) is hermitian, $\F$ is purely imaginary. Furthermore, $\F$ is symmetric:
$\F^{\a\b}=\F^{\b\a}$. 
Thus  we conclude from the above equation that $\F$ is identically zero: 
$
\F^{\a\b}=0.
$

The third equation from \eq{susyeq} leads to the expression (\ref{Bfield}) for $\B$.
Using the third equation from \eq{susyeq} and the fact that $\rho$'s anticommute, the first
equation from \eq{masterpr} can written as in \eq{jacobi}. 
  
In the supersymmetric case a closed expression for the star-product can be derived from the path integral. 
It reads
\bea
f\star g~(x,\theta) &=&  f(x,\t )~ \ex\LB {-i \hbar\ov 2}  \lde_{\a} \C^{\a i} \rde_i \RB \ex \LB{i \hbar\ov 2} \lde_i \B^{ij} \rde_j \RB \ex  \LB {i \hbar\ov 2} \lde_i  \C^{\a i} \rde_{\a} \RB \cdot \non
&&~~~ \ex\LB {\hbar^2\ov 12}  \C^{\a k} \de_{\a} \B^{ij} (\lde_i \rde_j\rde_k - \lde_i \lde_k \rde_j )\RB
g(x,\t).
\la{starproduct}
\eea
The Grassmann derivatives $\rde_{\a}$ and $\lde_{\a}$ in this equation act on the Grassmann variables as follows:
\be
\rde_1 \t^2 \t^1 = -\t^2,~~~~\rde_1 \t^1 \t^2 = \t^2,~~~~  \t^1 \t^2 \lde_1 = -\t^2,~~~~ \t^2 \t^1 \lde_1 = \t^2.
\la{conven}
\ee

We sketch the path integral derivation of \eq{starproduct}.  Schematically, we can write  the formula
(\ref{formula2}) 
as
\be
\langle f(X,\T)  ~\ex \LB  \B + \C  \RB ~g(X,\T) \rangle = \sum_{m,n} {1\ov m!} {1\ov n!}\langle 
f(X,\T)  ~ \B^m  \C^n~g(X,\T) \rangle ,
\la{symbolic}
\ee
where $\B = \B^{ij}\br_i \br_j$ and $\C = \C^{\a i} \bxi_{\a} \br_i$.
Since $\B^{ij}$ and $\C^{\a i}$ are $x$-independent, the $\br$'s in $\B$ from \eq{symbolic} can Wick-contract only with the
functions $f$ and $g$. The Wick contraction of $f$ and $\br_i$ gives a derivative of $f$: $\de_i f$.
Furthermore, if one of the $\br$'s from $\B$ contracts with $f$, then the remaining
$\br$ has to contract with $g$ to produce a non-vanishing result. 
This is  because  $\B^{ij}$ is anti-symmetric and so $\B^{ij} \de_i\de_j f$ vanishes. We will shortly show
that this type of contraction gives rise to the second exponential factor in \eq{starproduct}. 
 
The fate of $\bxi_{\a}$ and $\br_i$
in $\C$ can also be determined: $\br$ contracts with $f$ and $\bxi$ contracts with $g$, and vice versa. 
This type of contraction gives rise to the first and third exponential factors in \eq{starproduct}.
Note that if $\bxi$ and $\br$ contract with
the same function, then the corresponding contribution vanishes since the resulting Feynman integral vanishes:
\be
\int d\phi(a,z)\wedge d\phi(a,z) = 0,
\ee
where $a$ is a point on the real-axis, by the anticommutativity of the wedge product.

Since $\B^{ij}$ is linear in $\T$, it can also contract with $\bxi_{\a}$ in $\C$. The  three $\br$'s from $\B$ and $\C$ have to contract
with the functions $f$ and $g$. This type of contraction gives rise to the last exponential factor in \eq{starproduct}.

We now do some combinatorics to show that three different types of contractions just described indeed exponentiate 
into the formula (\ref{starproduct}). Let us pick $k$ $\B$'s and $k$ $\C$'s from \eq{symbolic}. There are 
$m!/(k! (m-k)!)$ times $n!/(k! (n-k)!)$ ways of doing that. There are $k!$ ways of contracting $k$ $\B$'s
with $k$ $\C$'s. Combining these combinatorial factors with the factorials from \eq{symbolic} we have
\be
{\B^{m-k}\ov (m-k)!} {\C^{n-k}\ov (n-k)!} {(\C\de \B)^k \ov k!}.
\ee
The fact that the Grassmann partial derivatives in \eq{starproduct} act as in \eq{conven} comes from the following
observation. Consider the correlator
\be
  \bigg\langle \cdots\bT \bT\cdots \bT  ~~ (\bxi_{\a} \C^{\a i} \br_i) \cdots \bigg\rangle
\ee 
where the $\bT$'s came from the Taylor expansion of $f(\t + \bT)$. In order to  Wick-contract a $\bT$ with
$\bxi$, we have to bring it next to the $\bxi$. Thus the Wick-contraction with the $\bxi_{\a}$ is effectively
equivalent to taking the derivative $\lde_{\a}$. 

The numerical coefficients in the exponents in \eq{starproduct} can be found using the integrals\cite{kont}
\be
\int d\phi(z_1,0)\wedge d\phi(z_1,1) = 2\pi^2
\la{integral1}
\ee
and
\be
\int d\phi(z_1,0)\wedge d\phi(z_1,1)\wedge d\phi(z_2,z_1)\wedge d\phi(z_2,1) = {4 \pi^4\ov 3}.
\la{integral2}
\ee
Each of the correlators (\ref{perturb})
 which appear as the coefficients of $\B^{ij} \de_i f \de_j g$, $\C^{\a i} \de_{\a} f\de_i g$ and $\C^{\a i} \de_{i} f\de_{\a} g$ is
proportional to the integral (\ref{integral1}). The correlator which appears as the coefficient of 
$\C^{\a k}\de_k \B^{ij} \de_{i} f \de_j \de_{\a}g$ is proportional to the integral (\ref{integral2}). The signs in the exponents
in \eq{starproduct} can be determined from the path integral as well.

\eq{starproduct} can, alternatively, be derived directly from the algebra (\ref{susyalg}) using the Baker-Hausdorff
formula:
\be
\e^A \e^B = \e^{A+B+{1\ov 2} [A,B]+{1\ov 12} ([A,[A,B]]+[[A,B],B])+\cdots}.
\ee
Thinking of the star-product as a matrix product and using the Baker-Hausdorff formula, we find:
\bea
&&\e^{i (p\cdot x + \eta\cdot\t)}\star \e^{i ({\tilde p}\cdot x +{\tilde \eta}\cdot\t)}= \non
&&= \ex \bigg( i(p+{\tilde p})\cdot x + i(\eta+{\tilde \eta})\cdot\t+
{i \hbar\ov 2}(\eta_{\a}\C^{\a i}{\tilde p}_i+p_i \C^{\a i}{\tilde \eta}_{\a}) \non
&&~~~~~~~- { i \hbar\ov 2} p_i {\tilde p}_j\B^{ij} -
{i\hbar^2\ov 12}  \C^{\a k} \de_{\a} \B^{ij} (p_i {\tilde p}_j {\tilde p}_k - p_i p_k {\tilde p}_j )\bigg).
\la{momprod}
\eea
Since
\bea
&&\e^{i p\cdot x} \lde_i = i p_i \e^{i p\cdot x},~~~\rde_i \e^{i {\tilde p}\cdot x}= i {\tilde p}_i \e^{i {\tilde p}\cdot x},\non
&&\e^{i \eta\cdot\t}\lde_{\a} = i \eta_{\a} \e^{i \eta\cdot\t},~~~\rde_{\a}\e^{i{\tilde \eta}\cdot \t} = -i{\tilde \eta_{\a}}\e^{i{\tilde \eta}\cdot \t} ,
\eea
the quadratic and cubic terms in momenta in the exponent in \eq{momprod} lead, after the substitutions 
\bea
&&p_k \rightarrow -i\lde_k,~~~~~~~{\tilde p}_k\rightarrow -i\rde_k, \non
&&\eta_{\a} \rightarrow -i\lde_{\a}, 
~~~~~~~{\tilde \eta}_{\a}\rightarrow i\rde_{\a},
\eea
to the star-product \eq{starproduct}.

\newsection{Summary}
Let us briefly summarize our results:

\noindent
$\bullet$
A path-integral formula (\ref{formula}) for the associative
star-product of functions of bosonic and fermionic coordinates $(x^i,\t^{\a})$ is proposed. $\t$ is
 taken to be a Majorana
spinor in $n$ spacetime dimensions.   

\noindent
$\bullet$
The BV action $S$ (\ref{action}) of the two-dimensional field theory depends on the background fields
$\B$, $\F$ and $\C$ from \eq{susyalgmod}. From the classical master equation $(S,S)_{\BV}=0$, we
derived the associativity conditions (\ref{master}) for the background fields. 

\noindent
$\bullet$ An operational formula (\ref{formula2}) is given for the star-product of functions of $x$ and $\t$
for general $x$- and $\t$-dependent background fields.

\noindent
$\bullet$  The condition of the supersymmetry 
 invariance of the action (\ref{action}) turned out to be very restrictive. The supersymmetric
backgrounds are given by the equations (\ref{Ffield}), (\ref{Bfield}) and (\ref{jacobi}). The supersymmetry
transformations (\ref{susy}) are $N=1$ or $N=2$ depending on the dimension $n$ of the spacetime.
An explicit formula (\ref{starproduct}) for the supersymmetric star-product is derived from the path-integral.

Ref.\cite{NSUSY} gives a deformation of $N=2$, $d=4$ superspace using symplectic-Majorana 
spinors. It turns out that a non-vanishing anticommutator 
$$
\{\t, \t\} \ne 0
$$
is compatible with the $N=2$ supersymmetry. 
With a minor modification of our action (\ref{action}) the case of the non-Majorana spinors can be dealt with.

\newsection{Acknowledgments}
The work of I.C. is supported by a Sherman Fairchild Prize Fellowship. The work of 
C.C. is supported in part by the DOE Grant DE-FG03-92-ER40701.

\end{document}